# Following wrong suggestions: self-blame in human and computer scenarios

Andrea Beretta[1,2] [0000-0001-8531-9325], Massimo Zancanaro[1,2] [0000-0002-1554-5703], Bruno Lepri[2][0000-0003-1275-2333]

[1] DIPSCO, University of Trento, Rovereto (TN), Italy

[2] Fondazione Bruno Kessler, Trento (TN), Italy

andrea.beretta@unitn.it

{zancana,lepri}@fbk.com

**Abstract.** This paper investigates the specific experience of following a suggestion by an intelligent machine that has a wrong outcome and the emotions people feel. By adopting a typical task employed in studies on decision-making, we presented participants with two scenarios in which they follow a suggestion and have a wrong outcome by either an expert human being or an intelligent machine. We found a significant decrease in the perceived responsibility on the wrong choice when the machine offers the suggestion. At present, few studies have investigated the negative emotions that could arise from a bad outcome after following the suggestion given by an intelligent system, and how to cope with the potential distrust that could affect the long-term use of the system and the cooperation. This preliminary research has implications in the study of cooperation and decision making with intelligent machines. Further research may address how to offer the suggestion in order to better cope with user's self-blame.

## 1. Introduction

Today intelligent systems are entering into our everyday life. These systems can help people to make more effective choices by generating predictions to the user in the form of advice and suggestions. Thus, in order to improve the design of these systems a greater understanding is needed on the conditions in which people emotionally deal with suggestions provided by an intelligent machine.

In this paper, we present an initial study investigating the emotions felt by people after following a wrong suggestion provided by a supposedly intelligent machine. We have focused on four specific emotions related to non-optimal decisions: *regret*, *disappointment*, *guilt*, and *perceived responsibility*.

Our results provides evidence that users' feelings of self-blame tend to be lower when they receive a wrong suggestion by a computer and the responses have a larger variation with respect to when the suggestion is provided by a human being.

## 2. Theoretical framework

Since the envision of Artificial Intelligence, technical progress has the intent of surpassing human performance and ability [1]. However, recently there is a growing interest in understanding the conditions for an effective cooperative relationship between human and computer agents as well as the possible biases of this cooperation.

In some recent studies, Logg [2, 3] has shown that people trust more a machine than other people when they need to make a decision in an objective context (e.g., they are looking for information). In other studies, in a subjective context (e.g., looking for book recommendations, looking for joke recommendations), people tend to rely more on other human beings [4, 5].

Machines can exceed human judgment in different ways. First, by exploiting the "wisdom of crowds", machines can surpass human accuracy in decisions even with simple algorithms (for example, averaging between the opinion of several individuals) [3, 6–8]. Second, algorithms trained with the same features used by human experts may exploit them more accurately [9]. Third, machine can automatically identify more predictive features than those commonly used by experts [10].

### 2.1 Emotions related with wrong decisions

Regret and disappointment are "*negative, cognitively determined emotions that we may experience when a situation would have been better if: (a) we had done something different (in case of regret); or (b) the state of the world had been different (in case of disappointment)*" [11]. They are the two most important emotions related with the decision process and both of them can be defined as counterfactual [12, 13]. Early theories have been studied by economists and investigated how the feeling of anticipated regret affects the decision process under uncertainty (e.g., [14–16] ). In psychological literature, the focus often is on how negative outcomes could intensify the experience of regret [17] and in how disappointment influences decision making [16, 18].

More specifically, regret is experienced when one could not obtain the expected goals, while disappointment arises when there is a goal abandonment [12]. Zeelenberg and colleagues argument that regret is related to a behavioral switch and decreases trust, whereas disappointment increases trust [19, 20]. Other studies argue that regret increases prosocial behaviors, while disappointment reduces them [19, 21, 22].

Another important aspect is how responsibility is related to these two emotions. Regret and disappointment differ from the perceived responsibility of the outcome: when a person feels more responsible for the bad outcome, regret is involved; while disappointment is involved when the person feels to not be responsible for the outcome [23, 24].

Particular attention should be paid to the difference between regret and guilt, both originated by negative outcomes related to a sense of self-responsibility. According to Zeelenberg and Breugelmans [25], regret and guilt are perceived as emotional outcomes of negative events and are related with a sense of self-agency and with the intention to change the event that happened. However, there are still not clear criteria to distinguish these two emotions while there is a general understanding that defines regret as a broader emotion than guilt [25, 26]. For example, Berndsen and colleagues [27] distinguish guilt from regret on the basis of interpersonal and more social factors related to guilt and intrapersonal factors more related to regret.

## 3. A user study on wrong decisions with intelligent machines

This study aims at investigating the emotional effects of a wrong decision taken after a suggestion by an intelligent machine with respect to when the wrong decision is taken after the suggestion from a human being. Our hypothesis is that when a suggestion is received by an intelligent machine (at least in an objective decision task), a wrong outcome may elicit less emotion directed toward the self, compared to the situation in which the suggestion is given by a human being. That is because an intelligent machine might be expected to provide more accurate suggestions [9] and the perceived agency is limited (unless the machine does have anthropomorphic features [28]).

Following a common practice in decision-making studies, we use scenarios rather than interactions with a real intelligent system in order to better control the conditions.

### 3.1 Hypotheses

Given the theoretical framework described above, and in particular following the results on decisions' outcomes of Logg [2,3], we expected that a suggestion provided by an intelligent machine (in a technical and relatively complicated decision task) decreases the possibility for the user to blame the advisor and therefore should produce a lower rating on counterfactual emotions and responsibility.

Therefore, after a wrong suggestion by an intelligent machine we postulate the following hypotheses:

- *Hypothesis on regret*: the user feels less regret than after a wrong suggestion given by a human being.

- *Hypothesis on disappointment*: the user feels more disappointment than after a wrong suggestion given by a human being.

- *Hypothesis on responsibility*: the user feels less responsible for the choice and for the bad outcome than after a wrong suggestion given by a human being.

- *Hypothesis on guilt*: the user experiences less guilt than after a wrong suggestion given by a human being.

### 3.2 Participants

Eighty-five participants were involved through the platform Amazon Mechanical Turk. Their age ranges from 18 to 66 years old (57 males, 24 females; mean=34 years; SD=9.55 years). The only requirement for participation was fluency in English.

### 3.3 Measures

In order to measure negative and counterfactual emotions, a questionnaire was built with 13 items obtained from validated scales:

- *Responsibility (self-blame)* is assessed by two items: the first measured the responsibility on the choice done, while the second one is about having done a bad purchase (i.e., bad outcome) [29].

- *Regret* is evaluated by two items aimed to assess the regret felt on the choice done and the regret felt for having a phone that does not meet the study participant's needs [29].

- *Disappointment* is assessed by two items aimed to evaluate the disappointment felt about having done a wrong choice and the disappointment for having a phone that does not meet the study participant's needs [29].

- *Perceived Guilt* is assessed by two items aimed to assess the perceived guilt on the choice done and the perceived guilt for having a phone that does not fulfill the study participants' needs [29].

The items are all 10-point Likert scales, ranging from strongly disagree (1) to strongly agree (10). Although Giorgetta [29] used 11-point scales, we eventually decided to not to have a central point.

Two items are added to control the experience of counterfactuals measuring the affective reaction and the dissatisfaction of the participant [30]. Finally, a manipulation check to prevent random answers is included (the item simply asks to select the "totally agree" score [31]).

### 3.4 Procedure

The experimental design is a between-subject with two conditions: (i) when a human being provides the suggestion (condition "Human") and (ii) when an intelligent machine provides the suggestion (condition "Computer").

At the beginning of the experiment, the participant received a short introduction to the task and s/he is asked to read and accept the informed consent and a short demographic questionnaire about age, gender and self-reported English fluency.

After that, the participant is automatically assigned to one of the two conditions and asked to read the first scenario about a purchase decision. The first scenario in "Computer" condition is as follows:

*"Imagine that you have to buy a new smartphone because yours has just stopped working. Even if you are not an expert, you go to the*

> *shop with the idea of buying XY10, because you think that is the model that best suits your preferences. Once in the shop, you enter your preferences in an **algorithm-based website** that suggests you buy smartphone WLx at the same price. Hence you decide to buy the suggested model WLx. Some time later, you realize that model XY10 would have been better for your needs, while the smartphone you have bought does not meet your expectations."*

The scenario in "Human" condition describes the same situation but it is different on the source of the advice:

> *"Imagine that you have to buy a new smartphone because yours has just stopped working. Even if you are not an expert, you go to the shop with the idea of buying XY10, because you think that is the model that best suits your preferences. Once in the shop, you explain your preferences to the **clerk** who suggests you buy smartphone WLx at the same price. Hence you decide to buy the suggested model WLx. Some time later, you realize that model XY10 would have been better for your needs, while the smartphone you have bought does not meet your expectations."*

Immediately after that, the questionnaire on negative and counterfactual emotions was administered for the first time. After the questionnaire, the participant was exposed to a second situation. The second scenario was about the same choice as before after a year and asked to check how dissatisfaction emotions change for the same decisional framework. Then, the participant was presented with the questionnaire on emotions for the second time.

At the end of the task, the participant was asked to recall the source of the suggestion received as a further attentional check.

Finally, the last step was about the debriefing of the participants on the actual aims of the study.

## 4.    Analysis

Before starting the analysis, the data has been checked for consistency. First, four participants were removed because they did not answer all the questions and another four participants were removed because they failed the consistency checks.

Then, 26 participants were removed because they showed inconsistent answers to the two control questions: "*I am sorry about what happened to me*" and "*I am satisfied about what happened to me*" in either the first or the second scenario.

Finally, 13 participants were removed because they took too long to complete the task. In order to identify these outliers, we adopted the Tukey's fences technique and we removed all the participants that took a time longer than 1.5 times the third quartile (4 minutes).

Eventually, 38 participants were retained for the study. The mean age was 32.5 years old (SD=7.39) ranging from 18 to 60 years old. Fifteen were women and 36 men. The average of the male sample has the mean equal to 31.44 years old (SD=6.57), and the female sample has mean 35.06 years old (SD=8.78).

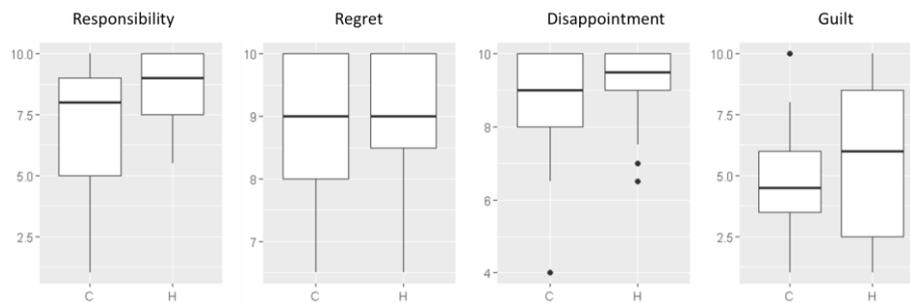

**Fig. 1.** The distribution of the negative and counter-factual emotions for the first scenario

The distributions of the emotions in the two conditions and in the two scenarios are shown in Figure 1 for the first scenario and in Figure 2, respectively.

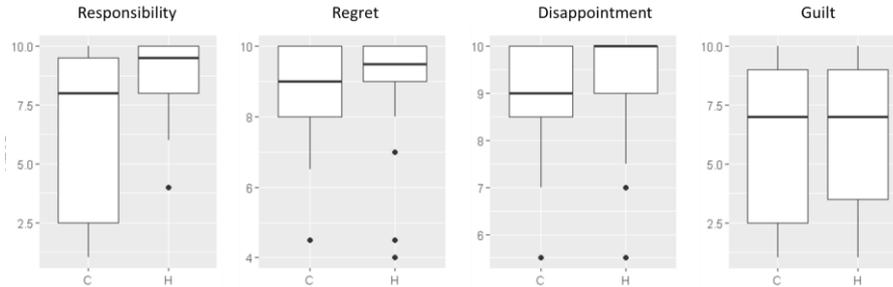

**Fig. 2.** The distribution of the negative and counterfactual emotions for the second scenario

For disappointment, regret and guilt there was no statistical difference in the two conditions in either scenarios. However, the responsibility on choice seems to change whether the source of suggestion was human or machine (using Kruskal-Wallis rank sum test, p = 0.09088 for the first scenario and p=0.04318 for the second scenario). The participants showed more self-blame when the suggestion came from a human being (mean=6.6 vs 8.4, sd=3.2 vs 1.6 in the first scenario and mean=6.4 vs 8.1, sd=3.6 vs 1.7 in the second scenario). Furthermore, the variance was much higher in the computer condition with respect to the human condition in both scenarios (Fligner-Killeen test of homogeneity of variances, p= 0.04051 and p= 0.00632, respectively).

## 5. Discussion

The results were not what we expected since we did not find any difference in regret, disappointment and guilt. Yet, our intuition was at least partially confirmed because the effect on responsibility was rather strong. Indeed, the effect on responsibility can be related to the antecedents of regret [32].

Even if further studies need to be carried out, from a Human-Computer Interaction (HCI) point of view, the decrease of the sense of responsibility when the suggestion comes from the computer may induce risky choices and it should be counterbalanced by appropriate design solutions.

The increase of variance in the responses when the suggestion comes from the computer may suggest that there might some cofounding variable that mediate this effect. This aspect should be better explored in further studies.

The study had some limitations. First, we may note that for all emotions and in both scenarios, the values tend to be quite high. This might be due to either a bias from recruiting the participants in a crowdsourcing service (as it might be apparent from the high number of participants that we had to exclude from the analysis) or from the scenarios that may have appeared confused or unnatural. Second, the variance is quite large for all the emotions and in both scenarios. Again, this might be due to the reasons above or it might depend by some conditions that we did not test (for example, different perceptions of the true intentions of the clerk while the computer might have look more neutral with respect to hidden intentions). Alternatively, personality traits (for example, the Locus of Control [33] of participants) might be a confounding variable in this case. Another possibility would be that participants might have transferred the responsibility of the wrong decision outcome on the source of the suggestion both perceived as experts (see for example the discussion in [24]).

## 6.    Conclusion and Future Work

In this paper, we presented a preliminary research aimed at understanding the possible differences in the acceptance of a suggestion from an intelligent machine with respect to a human being. We adopted the specific lens of analyzing negative and counterfactual emotions when the choice is eventually wrong.

Our study, in which we manipulated the outcome for controlling the experimental condition, found that participants' feeling of self-blame is lower when they receive a wrong suggestion by a computer rather than by a human being. This may suggest that decision making with computer advice may eventually induce risky choices.

This result needs to be confirmed by other studies, in particular in view of the lack of significance for the other negative and counterfactual emotions that are well-known related with sense of responsibility. In particular, it will be important to vary the scenarios and to control for possible cofounding variables (such as personality traits or attitude toward technology).

Still, we believe that our results are interesting for a better understanding of the relation between users and intelligent machines in a decision-making process and for a better design of this type of systems.